# Berezovsky number

Mikhail Simkin

and they that weave networks shall be confounded.
*Isaiah 19:9*

An important place in science has the so-called Erdős number [1]. It stems from the prolific mathematician Pál Erdős, who, in collaboration with 504 co-authors, proliferated 1416 scientific articles in the fields of number theory, probability theory and graph theory. For a concrete scientist, Erdős number determines his proximity to Erdős in the scientific collaboration space. Thus, for all 504 Erdős' co-authors this number equals 1. For co-authors of Erdős' co-authors (there is already 6,593 of those) this number equals 2. And so on: co-authors of people, who have Erdős number *n*, have Erdős number *n* + 1. The website of the American Mathematical Society has a tool [2], where after typing scientist's name you immediately get his Erdős number. It is very important for scientists to have small Erdős number. Apart from causing a well-deserved sense of accomplishment, it helps to win grants and Nobel prizes. In its importance in the life of a scientific animal, Erdős number is second only to Citation Index [3].

No doubt, that Pál Erdős is an interesting, important, and controversial figure. However, he is not the only one we can use to count numbers. Another worthy candidate is the corresponding member of the Russian Academy of Science Boris Berezovsky. He was sentenced in absentia to thirteen years in jail for fraud, embezzlement and money laundering. However, Berezovsky defenders argue that he is an honest mathematician with an Erdős number of 4. American Mathematical Society database holds 16 Berezovsky's articles [4]. He produced them with the help of 9 accomplices, all of whom are listed in Table 1.

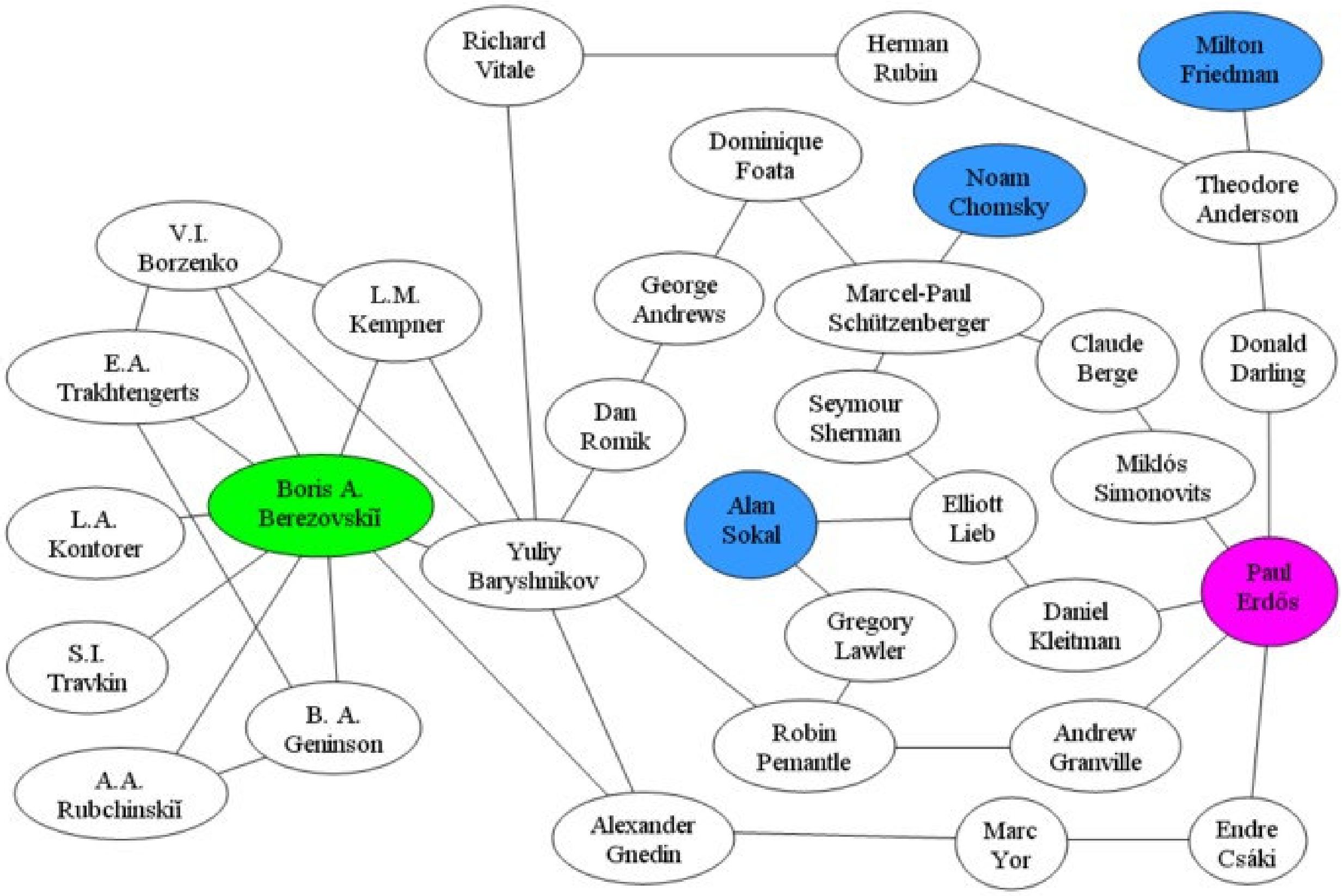

**Figure 1.** Berezovsky network. A link between names means that these people co-authored an article.

**Table1.** Berezovsky's closest accomplices (Berezovsky number = 1)

| | | | | |
|---|---|---|---|---|
| Baryshnikov, Yuliy. M. | Borzenko, V. I. | Kempner, L. M | Kontorer, L. A. | Gnedin, Alexander V. |
| Travkin, S. I. | Geninson, B. A. | Rubchinskii, A. A. | Trakhtengerts, E. A. | |

Berezovsky number is defined analogously to Erdős number. Thus, all 9 Berezovsky's co-authors get a number of 1. Co-authors of his co-authors (there is already 72 of those, and they are listed in Table 2) get a number of 2, and so on.

One can determine Berezovsky number using the same tool, which is used to compute Erdős number [2]. Just instead of "Erdős" you need to type "Berezovskii." It is important to spell his name the way I just did, since the database uses different spelling of his name than most of newspapers. I have computed Berezovsky number for few important people. It turned out that Alan Sokal is linked with Berezovsky and got a number of 4. Milton Friedman and Noam Chomsky are also connected with Berezovsky and got the numbers of 5 and 6 respectively. Figure 1 shows Berezovsky network uncovered in this investigation.

It would be dishonest to finish this article without revealing my own connection with the criminal world. My Berezovsky number is 6, and Figure 2 shows the chain linking me to Berezovsky.

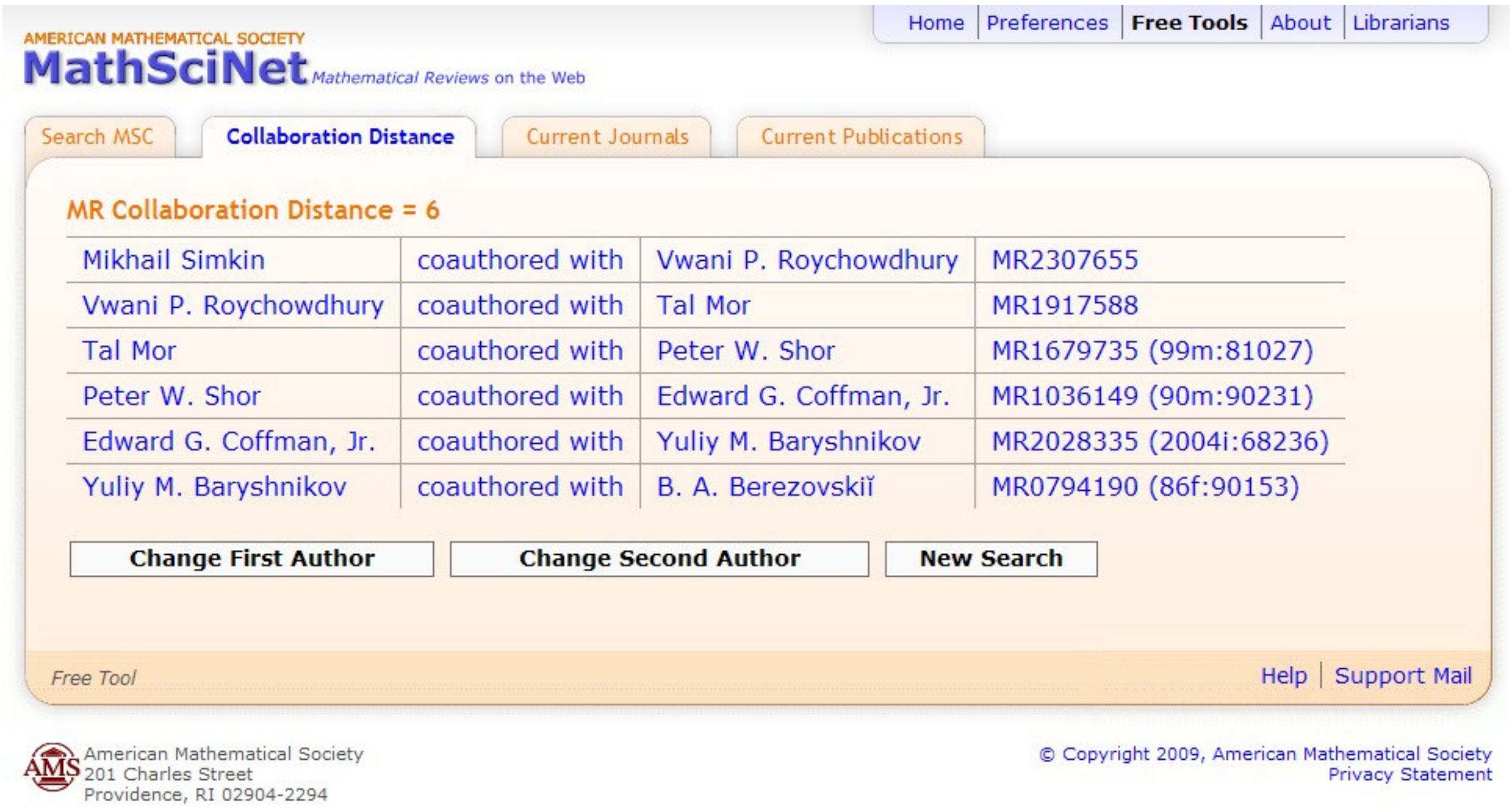

**Figure 2.** Connection of the present author with Berezovsky.

**Table 2.** Close Berezovsky's accomplices (Berezovsky number = 2).

| | | | | |
|---|---|---|---|---|
| Assaf, David | Babichev, Andrei V. | Barbour, Andrew D. | Belkina, M. V. | Bertoin, Jean |
| Bogachev, Leonid V. | Broido, Andrej | Bronshtein, I. I. | Chebotarev, Pavel Yu. | Chichilnisky, Graciela |
| Coffman, Edward G., Jr. | Derkach, B. T. | Dong, Rui | Dubov, Yu. A. | Eichelsbacher, Peter |
| Eisenberg, Bennett | Feng, Jing | Gaft, M. G. | Gendler, M. B. | Gritsyk, V. V. |
| Groppen, V. O. | Hansen, Ben B. | Itskovich, E. L. | Jelenkovic, Predrag R. | Kerov, Sergei Vasil'evich |
| Khalilov, A. I. | Khoroshavina, G. F. | Kolubai, S. K. | Krengel, Ulrich | Kuz'min, V. B. |
| Lazarevich, E. G. | Lebedev, V. G. | Lezina, Z. M. | Loginov, A. K. | Makarov, I. M. |
| Matakova, F. M. | Miretskiy, Denis I. | Mishchenko, V. A. | Momcilovic, Petar | Ol'shanskii, G. I. |
| Orlova, E. S. | Pitman, James W. | Podinovskii, Vladislav V. | Polyashuk, M. V. | Pronina, V. A. |
| Rozenblat, M. A. | Rubenstein, Daniel | Sakaguchi, Minoru | Sarychev, Andrei V. | Schreiber, Tomasz |
| Sergeev, V. I. | Shcherbakov, A. V. | Shershakov, V. M. | Shubina, M. V. | Shumei, A. S. |
| Shuraits, Yu. M. | Sokolov, V. B. | Stadje, Wolfgang | Stengle, Gilbert | Trishina, Elena |
| Tsodikova, Ya. Yu. | Val'kovskii, V. A. | Vinogradskaya, T. M. | Vitale, Richard A. | Volk, N. N. |
| Yakimets, V. N. | Yakovenko, Sergei | Yakubovich, Yu. V. | Yaralov, A. A. | Yor, Marc |
| Yukich, Joseph Elliott | Zharnitsky, Vadim | | | |